\documentstyle[twocolumn,epsf]{jpsj}
\def \virg{\;\;,}
\def \point{\;\,.}
\def \kf{k_{\rm F}}
\def \vf{v_{\rm F}}
\def \e{{\rm e }}
\def \i{{\rm i }}
\def \d{{\rm d}}
\def \td{t_{\rm d}}

\def \Vc{V_{\rm c}}
\def \V2c{V_{\rm 2c}}
\def \sgn{{\rm sgn}}

\def\ggs{\buildrel\textstyle > \over {\hbox{\raise0.2ex\hbox{$\sim$}}}}
\def\lls{\buildrel\textstyle < \over {\hbox{\raise0.2ex\hbox{$\sim$}}}}
\def\gsim{\,\lower0.75ex\hbox{$\ggs$}\,}
\def\lsim{\,\lower0.75ex\hbox{$\lls$}\,}

\def \SLP{\stackrel{\leftrightarrow}{\Pi}\!}
\def \SLU{\stackrel{\leftrightarrow}{U}\!}
\def \SLG{\stackrel{\leftrightarrow}{G}\!}
\def \SLH{\stackrel{\leftrightarrow}{H}\!}
\def\runtitle{
Role of Collective Mode for Optical Conductivity in 1/4-Filled SDW
  }
\def\runauthor
{
 Yuh {\sc Tomio} 
 and   
 Yoshikazu {\sc Suzumura} 
}

\renewcommand{\theequation}{\arabic{section}.\arabic{equation}}

\title{  
Role of Collective Mode for 
 Optical Conductivity  and Reflectivity \\
  in  Quarter-Filled Spin-Density-Wave State  
}

\author{
 Yuh {\sc Tomio}%
\footnote{E-mail: tomio@edu2.phys.nagoya-u.ac.jp}
   and Yoshikazu  {\sc Suzumura}%
\footnote{E-mail: e43428a@nucc.cc.nagoya-u.ac.jp}%
}

\inst{
Department of Physics, Nagoya University, Nagoya 464-8602 \\ 
}

\recdate{\hspace{20mm}}

\abst
{
Taking account of a collective mode relevant to charge fluctuation,  
  the optical conductivity   of  spin-density-wave  state
   has been examined    for an  extended   Hubbard model 
     with one-dimensional quarter-filled band. 
 We find  that,  within  the  random phase approximation, 
  the conductivity exhibits several peaks at the frequency 
   corresponding to the  excitation  energy 
    of the commensurate collective mode. 
  When  charge ordering  appears with increasing 
      inter-site repulsive   interactions,  
      the main  peak with  the lowest frequency  
         is reduced     and  the effective mass of the mode is enhanced 
       indicating the suppression 
       of the effect of the collective mode by  charge ordering. 
 It is also shown   that
 the reflectivity becomes large in  a wide  range of frequency 
    due to the huge dielectric constant induced by the 
  collective mode.  
}

\kword
{
optical conductivity, collective mode, 
spin-density-wave, charge ordering, reflectivity,
quarter-filled band, extended Hubbard model
  }

\begin{document}
\sloppy
\maketitle
\section{Introduction}
\setcounter{equation}{0}
 It has been well known that  collective modes 
  of the charge fluctuation  play   an important role to determine 
    the property of  density waves,  
  e.g.,   the pinning of the density wave  and 
     the dynamics such as the optical conductivity
  in    organic conductor.
\cite{Jerome,Gruner_rev,Quinlivan,Donovan,Gruner_Science}

  The  effect of the collective mode of  density waves 
    on the optical conductivity is  found in  the  pronounced  peak
        at zero frequency, which gives rise to   
       a sliding motion of the  density wave. 
   The collective mode  of the  charge-density-wave (CDW) state
      leads to the zero-frequency peak of the conductivity, 
       whose magnitude     depends  on the strength 
         of the coupling to phonon. 
 For the spin-density-wave (SDW) state, 
     the conductivity is determined only by  
  the collective mode where  the full weight is  given by the peak. 
 Such a fact is well known only for 
     the  incommensurate density wave.
 The optical conductivity for  incommensurate density waves 
 has been extensively examined since the work by Lee, Rice and Anderson
\cite{Lee74}
  who calculated  the conductivity by taking account of
    the collective mode  
      for both CDW and   SDW.
 The collective mode, which participates in   the sliding motion 
  of the CDW, has the effective mass enhanced by 
   the  electron-phonon interaction. 
 For large  effective mass,   
  the dominant contribution to  the  conductivity comes from  
    the single particle excitation.    
 On the other hand,   the conductivity of SDW state
    exhibits only the  zero frequency peak with the unrenormalized
    electron mass, i.e.,    
  the contribution from the single particle excitation 
  diminishes due to complete compensation 
   with that of the collective mode 
   and  the collective mode determines 
      the conductivity for all the frequencies.
The conductivity of SDW has been calculated 
  analytically  by Takada.
\cite{Takada84} 
 The numerical calculation of the conductivity with 
   the several electron-phonon   interactions 
       has been performed   by Fenton and Ares.
\cite{Fenton} 
Virosztek and Maki\cite{Maki88}
 examined  the pinning  effect 
 and the optical conductivity  by introducing   the  phase Hamiltonian. 

 The calculation of  the conductivity for 
  the commensurate case is  complicated 
  since  all the higher harmonics of the density wave 
  must be taken into account.
For understanding  the conductivity in the presence of 
   the commensurability and interaction,
 it is useful to calculate   a  sum rule showing  that 
  the summation of the conductivity   with respect to the frequency 
  is equal to the average of the kinetic energy.
\cite{Maldague77,Baeriswyl87} 
 The sum rule  depends  on the magnitude of the interaction
   except for  the limit of the weak coupling of
  the interaction as seen from  the incommensurate case. 
The exact numerical calculation
 for the  extended Hubbard model at quarter-filling 
 has  shown   the reduction  of the averaged value.
 \cite{Mila95}
This indicates an enhancement of effective mass 
 of density wave. 
 The optical conductivity of the Hubbard model 
  is further examined by introducing 
   the dimerization, which has an effect  of producing the charge gap.
\cite{Favand}
 However the calculation of  the optical conductivity 
   based on  the  mean field theory  is required 
       for the SDW state 
  since the  exact calculation of one-dimensional system leads to 
   the absence of the long range order.  

The SDW state of  organic   conductors
 has been studied extensively in terms of  the mean-field  theory. 
 In order to understand the experiment in   organic conductors  
where  SDW coexists  with  CDW,   
\cite{Pouget,Kagoshima} 
 several authors 
\cite{Seo,Kobayashi,TomioJPSJ}
 have examined the ground state of the extended Hubbard model 
 at quarter-filling.
 They have shown  such a coexistence 
 when   the inter-site repulsive interaction becomes 
 larger than the critical value.
Based on the mean field ground state, 
 the collective modes for charge fluctuation have been  
 calculated where the charge gap of the collective mode 
 vanishes at the onset of charge ordering.
\cite{Suzumura97,TomioLET} 
This comes from the second order phase transition between two kinds of 
 ground states. 
Thus 
 it is of interest to examine a role of such a low lying excitation 
 on the dynamical quantities.    
The optical conductivity at quarter-filling was  
 calculated  within  the single  particle excitation, 
\cite{Tajima}
and the effects of collective excitations on absorption spectra were  
 studied by applying the random phase approximation to the systems 
 with a finite size.    
\cite{Mori,Mori2}    
The purpose of the  present paper is to demonstrate such 
 a role of collective mode for   the optical conductivity 
 and the  reflectivity.
  Further we  examine  the effect of 
      charge ordering, which may originate in 
         inter-site repulsive interaction.

In \S2, formulation for the optical conductivity is given.
   Based on the mean-field ground state,   
  the random phase approximation  is applied to calculate   
   the response function for the electric current.  
In \S3, the optical conductivity is calculated numerically. 
 The effective mass is estimated from the main peak of the conductivity.
 In \S4, the effective mass of the collective mode is calculated 
    in terms of the phase Hamiltonian, which is derived from the 
     response function.  The result is compared with that of \S3. 
 Based on the optical conductivity, we examine the dielectric constant 
 and reflectivity   in \S 5.
 The plasma frequency 
 is calculated from the sum rule of the optical conductivity.  
The results is analyzed 
 in terms of the collective mode with low lying excitation. 
 In \S6, the effect of dimerization on the conductivity 
 and the effective mass is  briefly discussed.

\section{Formulation}
\setcounter{equation}{0}
We consider  a one-dimensional extended Hubbard model  given by  
   $H=H_0+H_{\rm int}$, where  $H_0$ and $H_{\rm int}$ are 
respectively defined by 
{\setcounter{enumi}{\value{equation}}
\addtocounter{enumi}{1}
\setcounter{equation}{0} 
\renewcommand{\theequation}{\arabic{section}.\theenumi\alph{equation}}
\begin{eqnarray}   
\label{H0}
	H_0   &=&  -\sum_{j=1}^{N}\sum_{\sigma=\uparrow, \downarrow}
         \left( t - (-1)^j \td \right)
         \left( C_{j\sigma}^\dagger  C_{j+1,\sigma} + {\rm h.c.} \right)
  \virg   \nonumber \\
   \\
\label{Hint}
  H_{\rm int} &=&   \sum_{j=1}^{N}
            \left(  U  n_{j \uparrow} n_{j \downarrow}
             + V    n_{j} n_{j+1}  
             + V_2  n_{j} n_{j+2} \right)  \point
\end{eqnarray}
\setcounter{equation}{\value{enumi}}}%
  In the Hamiltonian,   $C_{j\sigma}^{\dagger}$  
  denotes a creation  operator 
   of an electron at the $j$-th site with spin 
    $\sigma =(\uparrow, \downarrow)$, 
      $n_{j}=n_{j \uparrow}+n_{j \downarrow}$  and 
      $n_{j\sigma}=C_{j\sigma}^{\dagger}C_{j\sigma}$.  
   A periodic    boundary condition is taken, i.e.,  
     $C_{j+N,\sigma}^{\dagger}=C_{j \sigma}^{\dagger}$ 
       with the total number of lattice site $N$.  
 The quantity $t$  denotes   the transfer energy  
  and $\td$ corresponds  to the dimerization 
  where $t$ and the lattice constant are taken as unity. 
 Quantities  $U$, $V$ and $V_2$ are the coupling constants for 
  repulsive interactions  of  the on-site, the nearest-neighbor site and 
 the next-nearest-neighbor site.

 For a quarter-filled band,  
  where the Fermi wave number is given by $\kf =\pi/4$, 
   the mean-fields (MFs) 
  representing  SDW and CDW 
 are  written as
 ($m=0,1,2$ and 3, and    
 $\sgn (\sigma) = +(-)$ for $\sigma = \uparrow (\downarrow)$) 
{\setcounter{enumi}{\value{equation}}
\addtocounter{enumi}{1}
\setcounter{equation}{0} 
\renewcommand{\theequation}{\arabic{section}.\theenumi\alph{equation}}
\begin{eqnarray}   
   S_{mQ_0} &=& \frac{1}{N} \sum_{\sigma=\uparrow,\downarrow}  
                \sum_{-\pi < k \leq \pi } \sgn (\sigma)  
                \left<C_{k\sigma}^\dagger C_{k+mQ_0,\sigma}
                 \right>_{\rm MF}  \virg 
                                  \label{OPS} \nonumber\\   \\ 
   D_{mQ_0} &=& \frac{1}{N} \sum_{\sigma=\uparrow,\downarrow}  
                \sum_{-\pi < k \leq \pi }   
                \left<C_{k\sigma}^\dagger C_{k+mQ_0,\sigma}
                \right>_{\rm MF} \virg    
                                   \label{OPD}  
\end{eqnarray} 
\setcounter{equation}{\value{enumi}}}%
 where $Q_0 (\equiv 2\kf=\pi/2)$  
   induces density waves with a periodicity of four lattices.   
 The   $z$-axis is taken as   the quantized axis of the SDW ordered state.   
 The expression $\left <O \right>_{\rm MF}$ denotes an average of  
  $O$ taken  by the MF Hamiltonian,
\cite{TomioJPSJ}  
 which is expressed as 
\begin{eqnarray}   \label{HMF}
    H_{\rm MF} & = & \sum_{0 < k \leq Q_0} \sum_{\sigma} 
  \Psi^\dagger_{\sigma}(k)  \SLH_{\rm MF}^{\sigma}(k) \, \Psi_{\sigma}(k)
 \nonumber \\
        & & {} 
\hspace{-12mm}
       + 
       NU \Bigl[ - \frac{1}{16} -\frac{1}{2} \Bigl( 
       {|D_{Q_0}|}^2 \!-\! {|S_{Q_0}|}^2 \Bigr)  
     - \frac{1}{4} \Bigl( D_{2Q_0}^{2} \!-\! S_{2Q_0}^{2} 
       \Bigr) \Bigr]
\nonumber \\
 && {}  
\hspace{-12mm}
       +       
         NV  \Bigl(  -\frac{1}{4} + D^2_{2Q_0}  \Bigr) 
       + NV_2 \Bigl( -\frac{1}{4} + 2 {|D_{Q_0}|}^2 
       \!-\! D^2_{2Q_0} \Bigr) 
\virg \nonumber \\
\end{eqnarray} 
 with   
$\Psi^\dagger_{\sigma}(k)$ being the four component vector
 given by  
$
\Psi^\dagger_{\sigma}(k) = 
  (C^\dagger_{k \sigma}, C^\dagger_{k+Q_0, \sigma }, 
        C^\dagger_{k+2Q_0, \sigma}, C^\dagger_{k+3Q_0, \sigma} )
$.
  The MF Hamiltonian has the following 
   matrix elements.  
{\setcounter{enumi}{\value{equation}}
\addtocounter{enumi}{1}
\setcounter{equation}{0} 
\renewcommand{\theequation}{\arabic{section}.\theenumi\alph{equation}}
\begin{eqnarray}
\label{HMF1}
  \left( \SLH_{\rm MF}^{\sigma}(k) \right)_{mm} \!&=& 
   -2 t \cos ( k+mQ_0 )  
            + U/4 + V + V_2  \!\!\virg \nonumber \\
                                     \\
\label{HMF2}
  \left( \SLH_{\rm MF}^{\sigma}(k) \right)_{m+1,m} \!&=& 
   ( \frac{U}{2}\!-\!2V_2 ) D_{Q_0} 
    \!\!-\! \sgn(\sigma)\frac{U}{2}S_{Q_0}
\nonumber \\
 &=&  \left( \SLH_{\rm MF}^{\sigma}(k) \right)_{m,m+1}^{*}
\virg \\
\label{HMF3}
  \left( \SLH_{\rm MF}^{\sigma}(k) \right)_{m+2,m} \!&=& 
   ( \frac{U}{2}\!-\!2V\!+\!2V_2 ) D_{2Q_0} 
    \!\!-\! \sgn(\sigma)\frac{U}{2}S_{2Q_0}
\nonumber \\
 & & {} + 2 \i \td \sin (k+mQ_0) 
\point
\end{eqnarray}
\setcounter{equation}{\value{enumi}}}%
In the present case of quarter-filling, 
   MFs can be  rewritten as
\cite{TomioJPSJ} 
   $S_0=0$, $D_0=1/2$, $S_{Q_0}=S^*_{3Q_0} \equiv S_1 \e^{\i \theta}$, 
   $D_{Q_0}=D^*_{3Q_0} \equiv D_1 \e^{\i (\theta-\pi/2)}$, 
   $S_{2Q_0}=S^*_{2Q_0} \equiv S_2$ and  $D_{2Q_0}=D^*_{2Q_0} \equiv D_2$. 
Quantities  $S_1(> 0)$, $D_1 (\geq 0)$, $S_2$ and $D_2$    
  denote  amplitudes for 
2$\kf$~SDW, 2$\kf$~CDW, 4$\kf$~SDW and 4$\kf$~CDW, respectively.
The quantity $\theta$ denotes  a phase of SDW.  
The phase diagram of the ground state on the plane of $V$ and $V_2$ 
 has been shown explicitly in the previous paper.
\cite{Kobayashi,TomioJPSJ,TomioLET}   
 There are three kinds of ground state on the plane of $V$ and $V_2$
 where respective  order parameters are given by 
   $S_1$ in  the region (I),  
   $S_1$ and $D_2$ in  the region (II)  and 
   $S_1$, $S_2$ and $D_1$ in the region (III).  
 The region (II) (region (III)) is given  for large $V$  ($V_2$).

 The MF Hamiltonian, eq.~(\ref{HMF}), 
   is diagonalized with the eigenvalues 
   $E_{l}(k)$ ($l$=1,2,3 and 4) and eigenvector 
   $|lk\sigma \rangle$ 
     where   four energy bands are defined as 
$E_{1}(k) < E_{2}(k) < E_{3}(k) < E_{4}(k)$. 
These quantities are given by,  
{\setcounter{enumi}{\value{equation}}
\addtocounter{enumi}{1}
\setcounter{equation}{0} 
\renewcommand{\theequation}{\arabic{section}.\theenumi\alph{equation}}
\begin{eqnarray}
\label{eigenEQ}
 \SLH^{\sigma}_{\rm MF} (k) \; |lk\sigma\rangle &=& E_l(k) \, |lk\sigma\rangle
\virg \\
\label{eigenVEC}
 |lk\sigma\rangle_m &=& F_{lm\sigma}(k) 
\virg
\end{eqnarray}
\setcounter{equation}{\value{enumi}}}%
where $ F_{lm\sigma}(k)$ is the component of the eigenvector. 
  We note  that  only $E_1(k)$ is occupied
   due to quarter-filling. 
The effect of commensurability of quarter-filling 
 is sufficiently large in the present choice of $U=4$, 
 where the gap between $E_1(k)$ and $E_2(k)$ 
 is much larger than the band  width of $E_1(k)$.
 
%
\begin{figure}[tb]
\begin{center}
 \vspace{2mm}
 \leavevmode
 \epsfxsize=7.9cm\epsfbox{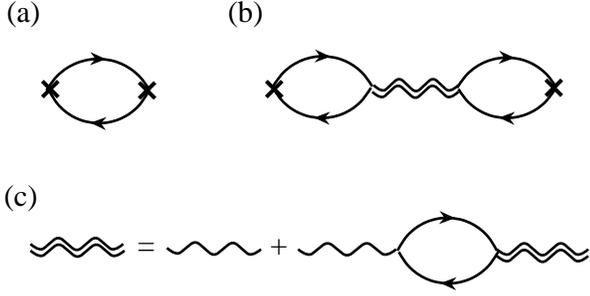}
 \vspace{-3mm}
\caption[]{
 Feynman diagram for the conductivity,
\cite{Lee74}
 where each  lines consist of the basis of four components.   
 The solid and wavy lines denote 
  the electron Green function and the effective interaction, respectively 
 and   the  cross denotes the current vertex.
 Respective diagrams denote 
(a) current-current response function for single particle excitation
  [$\sigma_0(\omega)$],  
 (b) current-current response function for collective mode 
  [$\sigma_c(\omega)$]
  and  
  (c) the effective interaction. 
 }
\end{center}
\end{figure} 
%
%
In order to examine the optical conductivity, 
 we use the following lattice version.
\cite{Baeriswyl87}
The current operator is   expressed as
\begin{eqnarray}
\label{J}
 J &=& \i e \sum_{j\sigma} \left( t - (-1)^j \td \right)
         \left( C_{j+1,\sigma}^\dagger  C_{j\sigma} 
              - C_{j\sigma}^\dagger  C_{j+1,\sigma} \right)
\nonumber \\
   &=& e \!\!\sum_{0 < k \leq Q_0,\sigma} \sum_{m=0}^{3}
\left( v_m (k) \; C^\dagger_{k+mQ_0,\sigma}C_{k+mQ_0,\sigma} 
\right. 
\nonumber \\
   & & \left. {}
 - \i  u_m (k) \; C^\dagger_{k+mQ_0,\sigma}C_{k+(m+2)Q_0,\sigma}
\right) 
\virg
\end{eqnarray}
where 
$v_m (k) = 2t \sin(k+mQ_0)$ and $u_m (k) = 2\td \cos(k+mQ_0)$. 
In terms of  eqs.~(\ref{J}), the Kubo formula for 
 the dynamical conductivity is given by 
\begin{equation}
\label{sigmadef}
 \sigma(\omega) + \i \sigma'(\omega)  = 
 \frac{\i}{N}  \int_0^{\infty} \d t \,
           \e^{\i \omega t}     
 \left< [ q(t), J(0)]\right>  
\virg
\end{equation} 
 where  $\sigma(\omega)$ and $\sigma'(\omega)$ are real and 
$q = e \sum_{j \sigma}    j \,C_{j \sigma}^{\dagger} C_{j \sigma}$. 
The optical conductivity is given by $\sigma(\omega)$. 
Within the random phase approximation, 
 the real part of eq.~(\ref{sigmadef}) is calculated as
\cite{Lee74} 
\begin{equation}
\label{sigma}
 \sigma(\omega)
 = \sigma_0 (\omega) + \sigma_c (\omega)
\virg
\end{equation} 
where $\sigma_0(\omega)$ and  $\sigma_c(\omega)$
 are respectively given by Fig.~1(a) and 1(b) 
as explained below. 
 For the convenience of numerical calculation, 
  we introduce a damping factor $\eta$ by   
  $\omega \rightarrow \omega + \i \eta$.   
  Thus a delta function corresponding to a pole 
   is replaced by a peak with a finite width. 

The first term of eq.~(\ref{sigma}) denotes the contribution from 
  single particle excitation, which is written as 
 [Fig.~1(a)] 
\begin{equation}
\label{sigma0}
 \sigma_0 (\omega) = {\rm Re} \left\{ \frac{2}{\i\omega} 
\left[ \Pi_{jj}^0(\omega) - \Pi_{jj}^0(0) \right] \right\}
\virg
\end{equation} 
where 
\begin{eqnarray}
\label{P0}
 \Pi_{jj}^0(\omega) &=& \frac{1}{2N} \int_0^\beta \d\tau
\left< T_\tau  J(\tau) J (0) \right>_{\rm MF} \e^{\i\omega_n \tau} 
\bigr|_{\i\omega_n \rightarrow \omega + \i \eta}
\nonumber \\
  &=&  -\frac{e^2}{2N} \!\!\sum_{0 < k \leq Q_0,\sigma}
                          \sum_{m_1,m_2}\sum_{l,l'}
\frac{f(E_{l'}(k)) \!-\! f(E_{l}(k))}
{\omega \!+\! \i \eta \!+\! E_{l'}(k) \!-\! E_{l}(k)} 
\nonumber \\
  & & {} 
 \times \left[ v_{m_1}(k) F_{lm_1\sigma}(k) 
           -\i u_{m_1}(k) F_{l,m_1+2,\sigma}(k) \right]
\nonumber \\
  & & {} 
 \times \left[ v_{m_2}(k) F_{lm_2\sigma}^*(k) 
           +\i u_{m_2}(k) F_{l,m_2+2,\sigma}^*(k) \right]
\nonumber \\
  & & {} \times F_{l'm_2\sigma}(k) F_{l'm_1\sigma}^* (k)
\virg
\end{eqnarray}
 and $T_{\tau}$ is the ordering operator for the imaginary time $\tau$. 
 The quantity $\beta^{-1}$ denotes a temperature with the Boltzmann 
 constant taken as  unity and  
$\omega_n=2\pi n \beta^{-1}$ with integer $n$.   
 The function $f(x)$ denotes a Fermi distribution 
 function given by  $[\e^{(x -\mu)\beta} + 1]^{-1}$
  where the chemical potential is given by  
  $\mu = (E_1(Q_0)+E_2(Q_0))/2$ in the limit of zero temperature.   
In deriving eq.~(\ref{P0}), we used a single particle Green function 
(the solid line in Fig.~1), written as 
$ 
(\SLG^{\sigma} (k, \i \omega_\nu))^{-1} 
 =  \i \omega_\nu - \SLH^{\sigma}_{\rm MF} (k) + \mu 
$
where $\omega_\nu$ is the Matsubara frequency for Fermion.
 
The second term of eq.~(\ref{sigma}) denotes  the  contribution from 
 the collective mode expressed as  
\begin{equation}
\label{sigmac}
 \sigma_c (\omega) = {\rm Re} \left\{ \frac{2}{\i\omega} 
             \Pi_{jj}^c(\omega) \right\}
\virg
\end{equation}
where 
\begin{eqnarray}
\label{Pc}
 \Pi_{jj}^c(\omega) &=& \frac{1}{2N} \int_0^\beta \d\tau
\left< T_\tau  J(\tau) J(0) \right> \e^{\i\omega_n \tau} 
\bigr|_{\i\omega_n \rightarrow \omega + \i \eta}
\nonumber \\
  &=& 
      \hat{\Pi}_{ju}(\omega) 
    \frac{\SLU(0)}{1-\SLU(0) \SLP(0,\omega)} 
      \hat{\Pi}_{uj}(\omega)
\point
\end{eqnarray}
 The matrix,  
$
 \SLU(q)/(1-\SLU(q) \SLP(q,\omega))
$, denotes an effective interaction (Fig.~1(c)). 
 The quantity,   $\SLU(q)$,   has only the  diagonal element given by 
$
( \SLU(q) )_{ii'} 
 = \delta_{ii'} U_i(q)
$, 
$(i=1,3,0,2)$ 
 where  
$ U_1(q) =  U_3(q) = U$,
 $U_0(q) =  - (U + 4V \cos q + 4V_2 \cos 2q )$ and 
 $U_2(q) =  - (U - 4V \cos q + 4V_2 \cos 2q )$.
In eq.~(\ref{Pc}), respective vector and matrix are written as  
\begin{eqnarray}
\label{Pju}
 \hat{\Pi}_{ju}(\omega) &=& 
\left(
\Pi^{01}_{jz}(\omega),\Pi^{03}_{jz}(\omega),
\Pi^{00}_{jd}(\omega),\Pi^{02}_{jd}(\omega)
\right) 
\!\!\!\virg \\
\label{Puj}
 \hat{\Pi}_{uj}(\omega) &=&
\left(
\Pi^{10}_{zj}(\omega), \Pi^{30}_{zj}(\omega), 
\Pi^{00}_{dj}(\omega), \Pi^{20}_{dj}(\omega)
\right)^{t}
\!\!\!\!\virg \\
\label{Pi}
\SLP(q,\omega) &=& 
\nonumber \\
 & & \hspace{-15mm} \left(
\begin{array}{cccc}
\Pi^{11}_{zz}(q,\omega) & \Pi^{13}_{zz}(q,\omega) & 
\Pi^{10}_{zd}(q,\omega) & \Pi^{12}_{zd}(q,\omega) \\ 
\Pi^{31}_{zz}(q,\omega) & \Pi^{33}_{zz}(q,\omega) & 
\Pi^{30}_{zd}(q,\omega) & \Pi^{32}_{zd}(q,\omega) \\ 
\Pi^{01}_{dz}(q,\omega) & \Pi^{03}_{dz}(q,\omega) & 
\Pi^{00}_{dd}(q,\omega) & \Pi^{02}_{dd}(q,\omega) \\ 
\Pi^{21}_{dz}(q,\omega) & \Pi^{23}_{dz}(q,\omega) & 
\Pi^{20}_{dd}(q,\omega) & \Pi^{22}_{dd}(q,\omega) 
\end{array}
\right)  \virg
\nonumber \\ 
\end{eqnarray}
The elements of eqs.~(\ref{Pju})-(\ref{Pi}) are 
 calculated as  
\begin{eqnarray}
\label{Pjz} 
  \Pi_{jz}^{0m}(\omega) &=& \frac{1}{2N} \int_0^\beta \!\d\tau 
\left<
 T_{\tau} J (\tau) S^\dagger_{zm}(0)
\right>_{\rm MF} \e^{\i\omega_n \tau}
\bigr|_{\i\omega_n \rightarrow \omega + \i \eta}
\nonumber \\
 &=& 
-\frac{e}{2N} \!\!\sum_{0 < k \leq Q_0,\sigma}
                          \sum_{m_1,m_2}\!\sum_{l,l'}
\frac{f(E_{l'}(k)) \!-\! f(E_{l}(k))}
{\omega \!+\! \i \eta \!+\! E_{l'}(k) \!-\! E_{l}(k)} 
\nonumber \\ 
 & & {} \times 
         \left[ v_{m_1}(k) F_{lm_1\sigma}(k) 
           -\i u_{m_1}(k) F_{l,m_1+2,\sigma}(k) \right]
\nonumber \\
  & & {} \times  
 F_{l, m+m_2, \sigma}^* (k) F_{l'm_2\sigma}(k) F_{l'm_1\sigma}^* (k)
 \;\sgn(\sigma) \virg
\nonumber \\ \\
\label{Pzj} 
  \Pi_{zj}^{m0}(\omega) &=& \frac{1}{2N} \int_0^\beta \!\d\tau 
\left<
 T_{\tau} S_{zm}(\tau,0) J 
\right>_{\rm MF} \e^{\i\omega_n \tau}
\bigr|_{\i\omega_n \rightarrow \omega + \i \eta}
\nonumber \\
 &=& 
- \frac{e}{2N} \!\!\sum_{0 < k \leq Q_0,\sigma}
                          \sum_{m_1,m_2}\!\sum_{l,l'}
\frac{f(E_{l'}(k)) \!-\! f(E_{l}(k))}
{\omega \!+\! \i \eta \!+\! E_{l'}(k) \!-\! E_{l}(k)} 
\nonumber \\ 
 & & {} \times 
          \left[ v_{m_2}(k) F^*_{lm_2\sigma}(k) 
           +\i u_{m_2}(k) F^*_{l,m_2+2,\sigma}(k) \right]
\nonumber \\
  & & {} \times
 F_{l, m+m_1, \sigma} (k) F_{l'm_2\sigma}(k) F_{l'm_1\sigma}^* (k)
 \;\sgn(\sigma) \virg
\nonumber \\ 
\end{eqnarray}
and 
\begin{eqnarray}
\label{Pzd} 
  \Pi_{zd}^{mm'}\!(q,\omega) &=&  \Pi_{dz}^{mm'}\!(q,\omega) 
\nonumber \\
 & & \hspace{-20mm} =
 \frac{1}{2N} \int_0^\beta \!\d\tau 
\left<
 T_{\tau} S_{zm}(\tau,q) D_{m'}^\dagger(q)
\right>_{\rm MF} \e^{\i\omega_n \tau}
\bigr|_{\i\omega_n \rightarrow \omega + \i \eta}
\nonumber \\
& &  \hspace{-20mm} =
- \frac{1}{2N} \!\!\sum_{0 < k \leq Q_0,\sigma}
                          \sum_{m_1,m_2}\!\sum_{l,l'}
\frac{f(E_{l'}(k)) \!-\! f(E_{l}(k+q))}
{\omega \!+\! \i \eta \!+\! E_{l'}(k) \!-\! E_{l}(k+q)} 
\nonumber \\
  & &  {} \hspace{-15mm} \times  
 F_{l, m+m_1, \sigma} (k+q) F_{l, m'+m_2, \sigma}^* (k+q)
\nonumber \\
  & & {} \hspace{-15mm} \times  
 F_{l'm_2 \sigma}(k) F_{l'm_1 \sigma}^* (k)
 \;\sgn(\sigma) 
\virg
\end{eqnarray}
where 
$S_{zm}(q)=\sum_{k\sigma}C_{k\sigma}^\dagger C_{k+q+mQ_0,\sigma}\sgn(\sigma)$
and 
$D_{m}(q)=\sum_{k\sigma}C_{k\sigma}^\dagger C_{k+q+mQ_0,\sigma}$.
Quantities  
$\Pi^{0m}_{jd}(\omega)$,  $\Pi^{m0}_{dj}(\omega)$ 
and $\Pi_{zz}^{mm'}\!(q,\omega)(=\Pi_{dd}^{mm'}\!(q,\omega))$
 are given by  eqs.~(\ref{Pjz}), (\ref{Pzj}) and  (\ref{Pzd}) 
without $\sgn(\sigma)$, respectively.

The spectrum of the collective mode  is obtained from 
\begin{equation}
 \label{pole}
  \det \left( 1 - \SLU(q)\SLP (q,\omega(q)) \right) = 0 
\virg
\end{equation} 
where the basis vector is given by
\cite{Suzumura_JPSJ95} 
\begin{equation}
 \label{base}
 \Phi(q)=  (S_{z1}(q),S_{z3}(q),D_0(q),D_{2}(q))^t \point
\end{equation} 
 The solution of eq.~(\ref{pole}) consists of several poles, i.e., 
  isolated solution.
 We define  $\omega_c(q)$ as the pole with the lowest energy 
   which corresponds to   the  mode 
   leading  to the  sliding motion of  density wave in the limit of weak 
 interaction (or  incommensurate case 
\cite{Lee74}). 
In addition to  $\omega_c(0)$ expressing    the charge fluctuation, 
 some of  other  poles with higher energy may also  
   give rise to  the peak of the optical conductivity 
   if they describe  the  charge fluctuation. 
 In the next section, 
  these mode are examined 
     by  rewriting  the basis of  eq.~(\ref{base}).  
 In eq.~(\ref{pole}), there are also three continua 
  due to quarter-filled band,  which contribute to 
  the conductivity.

 For calculating  the effective mass, we note   
    sum rules 
   for $\sigma (\omega)$ and $\sigma_c(\omega)$  
    of  eq. (\ref{sigma}) expressed as  
\begin{equation} \label{sum1}
  \int_0^\infty \hspace{-3mm}\d\omega \:\sigma(\omega) =
 - \frac{\pi e^2}{2N} \left< H_0 \right>_{\rm MF} 
 \equiv \frac{m}{m_c} \sqrt{2} \:e^2
\virg  
\end{equation} 
\begin{equation} \label{sum2}
  \int_0^\infty \hspace{-3mm}\d\omega \:\sigma_c (\omega) = 0
\virg  
\end{equation} 
 where $m (= \pi/(4 \sqrt{2}))$ is the band mass.  
 In these equations, 
 the former  is known analytically
\cite{Maldague77,Baeriswyl87} 
and the latter  is  verified numerically in the present paper.
 The mass, $m_c$, which is reduced to  $m$ for the incommensurate case, 
  is renormalized by interactions.  
The effective mass of the collective mode can be 
  obtained from the weight  of the main peak of the conductivity.  

\section{Optical Conductivity}
\setcounter{equation}{0}

 The frequency dependence of optical conductivity 
  is examined  to comprehend  the role of the collective mode 
   of the commensurate density wave at  quarter-filling. 
 Parameters $U$,  $V$ and $V_2$ are chosen  as  intermediate coupling 
   e.g,  $U$ = 3,  4 and 5. 
  We take  $\td = 0$ in this section and  
  discuss the effect of $\td \not= 0$ (i.e., dimerization) 
    in the final section.

Since  the fundamental contribution to the optical conductivity 
  comes from  the single particle excitation,
  we  calculate 
  the density of states  defined by 
\begin{equation} \label{dos}
 D(\omega) =
 - \frac{2}{ \pi N} \sum _{0<k<Q_0} {\rm Im} \{ {\rm Tr}  
  \SLG^{\sigma} (k, \omega + {\rm i}0 -\mu)
   \}
  \virg 
\end{equation} 
i.e.,
$
D(\omega)= 
   2/(\pi N) {\rm lim}_{\delta  \rightarrow  +0}
\sum_{k, \sigma} \sum_l
          \delta /((\omega-E_l(k))^2+\delta^2) 
$
where 
$
\int \d \omega D(\omega ) = 2
$. 
 In  Fig.~2,
  the density of states, $D(\omega)$ is shown for $U$ = 4, where   
    three  choices of parameters of $V$ and $V_2$  
      correspond  to region (I), region (II) and region (III), 
 respectively. 
  The occupied states are shown by the shadow, the area of which is  1/2.  
 At  quarter filling,  
   density of states are divided into four parts,
  corresponding to $E_1(k)$, $E_2(k)$, $E_3(k)$ and $E_4(k)$. 
 There exists  a  symmetry with respect to $\omega =(E_2(Q_0)+E_3(Q_0))/2$
  only for  $V = V_2 = 0$. 
 The   gap  between the first band 
   ($E_1(k)$) and the second one ($E_2(k)$)  
      increases with increasing $V$ and/or $V_2$ 
   while the second gap is  nonmonotonical due to 
     vanishing at $V=U/4$ in the region (II) 
       as seen from eqs. (\ref{HMF2}) and (\ref{HMF3}) with $\theta = 0$.

%
\begin{figure}[tb]
\begin{center}
 \vspace{2mm}
 \leavevmode
 \epsfxsize=7.5cm\epsfbox{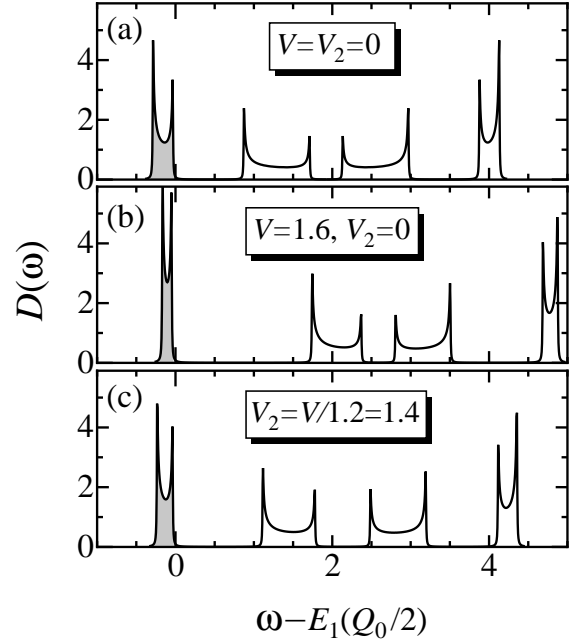}
 \vspace{-3mm}
\caption[]{
The density of states for $U=4$, $\td=0$ and with  the fixed 
 (a)  $V=V_2=0$ (region (I)), (b) $V=1.6$ and $V_2=0$ (region (II)), 
 and (c) $V_2=V/1.2=1.4$ (region (III)). 
 }
\end{center}
\end{figure} 
%
%
\begin{figure}[tb]
\begin{center}
 \vspace{2mm}
 \leavevmode
 \epsfxsize=7.9cm\epsfbox{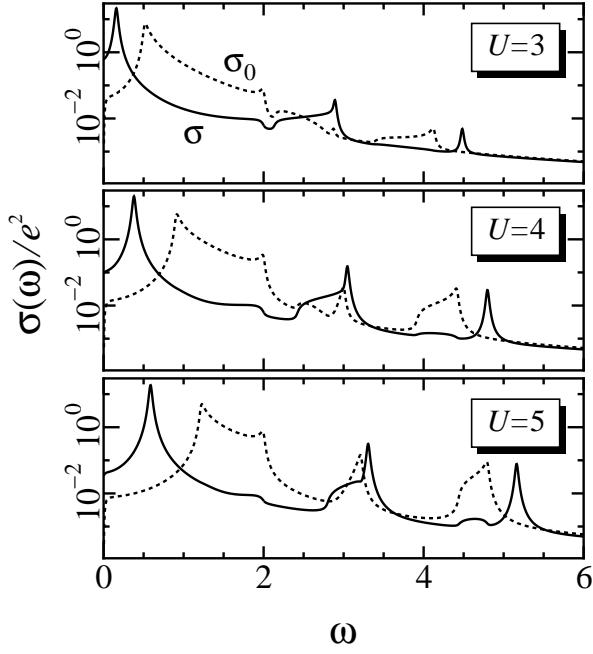}
 \vspace{-3mm}
\caption[]{
Optical conductivity  $\sigma (\omega)$ is shown by the solid curve
for $U=3$, 4 and  5  with $V=V_2=0$, $\td=0$  and $\eta=0.02$ 
 where $\sigma (\omega)  = \sigma_0(\omega)+\sigma_c(\omega)  $
  and  the dotted curve denotes  $\sigma_0$, 
   the contribution from  single particle excitation.
 }
\end{center}
\end{figure} 
%
Now we calculate the optical conductivity. 
Figure  3 shows  the conductivity  for the Hubbard model 
  with $U$ = 3, 4 and 5, where dotted curve and solid curve 
   correspond to $\sigma_0(\omega)$ and  $\sigma(\omega)$, respectively. 
 The dotted curve shows the three continua, 
  which correspond to  the excitation    from the filled 
 (i.e., first) band to 
  the second  band, the third one and the fourth one. 
 The solid curve exhibits  three sharp peaks, which  have the following 
  relation to 
    the spectrum  of  collective modes at $q=0$.  
 For $U$=4, the peaks are located at  
   $\omega \simeq$  0.4, 3.0 and  4.8 while 
     eq.~(\ref{pole}) with $q=0$      gives 
  five  poles at $\omega$ =  0.38, 2.24, 3.05, 3.88 and  4.80
   respectively 
     with three  continua  
        $ 0.91 < \omega < 2.00$, 
          $2.42 < \omega < 3.01$  and 
         $ 3.91 < \omega < 4.42 $.
  It is found that the peaks of $\sigma (\omega)$  correspond 
     to the first, third and fifth poles. 
    The contribution from the second  and fourth  poles 
          is absent within the numerical accuracy of 
              the present calculation. 
 For $U$ = 3, the second and third  peaks  of  $\sigma (\omega)$ 
   are reduced.  This  indicates  a fact that 
  $\sigma (\omega)$ in  the limit of small $U$ 
    is reduced to that of the incommensurate case 
   where all the contribution of $\sigma (\omega)$ is given by 
    the  first peak with  $\omega \rightarrow 0$ 
  and the contribution from the single particle 
   (i.e. $\sigma_0(\omega)$) compensates   completely with that of 
     $\sigma_c(\omega)$. 
  With increasing $U$, 
    the weight for the lowest peak decreases and 
      the second and the third peaks increase 
       together with  the continuum   
       especially in the frequency region   just below the second peak.  

%
\begin{figure}[tb]
\begin{center}
 \vspace{2mm}
 \leavevmode 
\epsfxsize=7.9cm\epsfbox{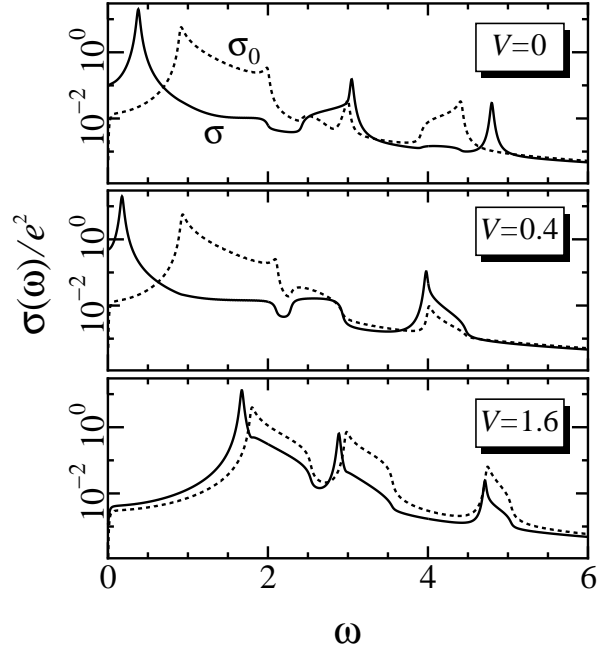}
 \vspace{-3mm}
\caption[]{
Optical conductivity  $\sigma (\omega)$ is shown by the solid curve
for $V=0$, 0.4 and 1.6  with $U=4$, $V_2=0$, $\td=0$  and $\eta=0.02$ 
 where the notations are the same as Fig.~3.
}
\end{center}
\end{figure}
%
%
\begin{figure}[tb]
\begin{center}
 \vspace{2mm}
 \leavevmode
 \epsfxsize=7.2cm\epsfbox{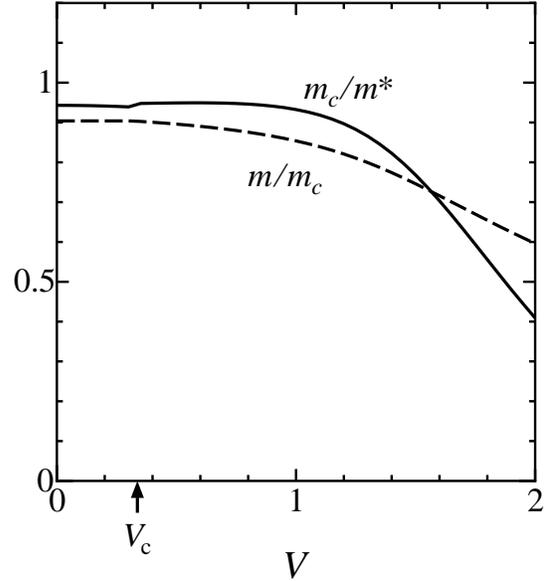}
 \vspace{-3mm}
\caption[]{
 $V$ dependence of $m/m_c$ (dashed curve) and $m_c/m^*$ (solid curve) 
    for $U=4$,  $V_2=0$ and $\td=0$.  
 The ground state in  the region (I) [region (II)] is obtained  
 for $ V < \Vc(\simeq 0.34)$   ( $ V > \Vc$). 
}
\end{center}
\end{figure}
%
The effect of $V$ on $\sigma(\omega)$ is examined in Fig.~4 
 for  $U$=4 and $V_2=0$ 
 where $\sigma(\omega)$  
 with $V= 0.4$ and 1.6     
  are compared with that of $V = 0$.
 The conductivity of the upper panel is 
  the same as that of $U=4$ in  Fig.~3.   
 The  region (II)  is obtained    for $V >  \Vc (=  0.34) $ 
  while  the  region (I) is obtained    for $V <  \Vc $. 
 For $V = 0.4$ (the middle panel),  
 the peaks   are located at 
    $\omega \simeq  0.2$ and  4.0  
 while   eq.~(\ref{pole}) with $q=0$      gives 
  four   poles at 
$\omega$ = 
 0.18, 2.21, 3.98 and  4.65
   respectively 
     with three  continua  in the regions 
        $  0.92 < \omega < 2.11$, 
           $2.28 < \omega < 2.93 $  and 
         $ 4.00 < \omega < 4.48 $.
  Thus, the peaks of $\sigma (\omega)$  correspond  
    to the first and  third  poles. 
  The effect of $\sigma_c$ is small
    for intermediate  frequencies. 
 This may come from the fact that the gap between 
 the second band and the third band (i.e., $E_2(0)$ and $E_3(0)$)
 vanishes at $V = U/4$ in the region (II) as seen 
  from eq.~(\ref{HMF3}) and $\theta =0$.  
 For $V = 1.6$ (the lower panel), 
   the peaks are located at  
   $\omega \simeq $ 
  1.7, 2.9 and  4.7
  while 
     eq.~(\ref{pole}) with $q=0$      gives 
  five  poles at 
$\omega$ = 
 1.67, 1.70, 2.89, 2.96 and  4.71
   respectively 
     with three  continua 
        $1.79 < \omega < 2.54$, 
           $2.97 < \omega < 3.56 $  and 
         $4.74 < \omega < 5.04$.
  Thus, the peaks of $\sigma (\omega)$  correspond 
    to  first, third and fifth   poles. 
  It is found that, within the numerical accuracy,   
   the second  and fourth  poles of  the collective mode 
    do not give the peak of the conductivity. 
 With increasing $V$,  the weight of the first peak is reduced.
For large $V$, 
  the compensation of the continuum by  $\sigma_c(\omega)$   is reduced
  and the continuum of $\sigma (\omega)$  becomes noticeable.

Here we  examine the total weight of the conductivity.
In Fig.~5,
 the quantity $m/m_c$ defined by eq.~(\ref{sum1}) as the function 
 of $V$ is shown by the  dashed curve 
 where  $m_c$ increases   with increasing  $V$ 
  due to narrowing of the filled band.     
Note that $m/m_c=1$ in the limit of small $U$ and $V$. 
Here we examine the effect of $V$ on the weight of 
 the main (i.e., first) peak of  $\sigma (\omega)$. 
Since the finite imaginary part $\eta$ is introduced 
 by treating $\omega \rightarrow \omega + \i \eta$,
 we calculate the weight using a method given  by  
\begin{equation} \label{sum3}
  \int_0^{\omega_0} \hspace{-3mm}\d\omega \:\sigma_c (\omega) \equiv
 \frac{m}{m^*} \sqrt{2} \:e^2
\virg  
\end{equation} 
where $\omega_0$ denotes the lowest value of $\omega$
  satisfying    $\sigma_c(\omega) = 0$, i.e., 
 the solid curve and the dotted curve in Fig.~4 cross each other  
  at $\omega = \omega_0$.  
 The quantity $m^*$ may be regarded as the effective mass
 for the collective mode with  the first (lowest) pole.
 In Fig.~5, 
 the quantity $m_c/m^*$, which denotes a fraction of   
 the collective mode is shown by the solid curve.
 With increasing $V$, the quantity $m_c/m^*$ decreases 
 indicating the decrease of the  contribution 
  from collective modes.
 There is a jump of $m_c/m^*$ at $V= \Vc$. 
 In  the limit of the weak coupling, one obtains   $m_c/m^* = 1$ 
   showing that the conductivity is determined only by a collective mode 
     with $\omega \rightarrow 0$.   

%
\begin{figure}[tb]
\begin{center}
 \vspace{2mm}
 \leavevmode
 \epsfxsize=7.9cm\epsfbox{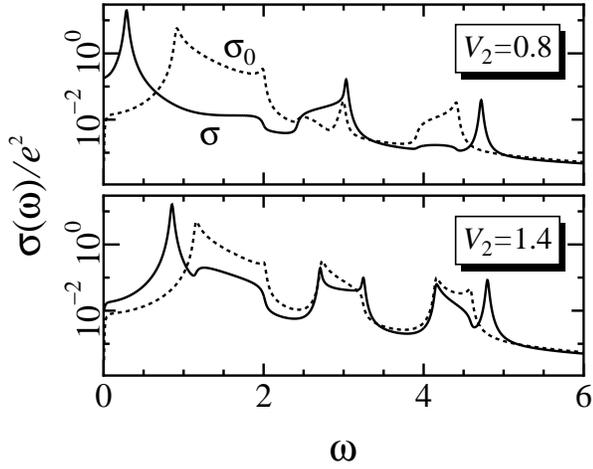}
 \vspace{-3mm}
\caption[]{
Optical conductivity  $\sigma (\omega)$ is shown by the solid curve
for $V_2 = 0.8$  and 1.4  with $U=4$, $V=1.2 V_2$, $\td=0$ 
  and $\eta=0.02$ 
 where the notations are the same as Fig.~3.
}
\end{center}
\end{figure}
%
The effect of $V_2$  on the conductivity is examined in Fig.~6 
    with $U =4$ and $V/V_2 =1.2$.
 Note that the parameter with  $V_2=0.8$ ($V_2=1.4$) corresponds 
      to  the region (I) [region (III)] as seen from a fact that  
 the boundary between region (I) and region (II) for $U =4$  is given by 
\cite{TomioJPSJ}
 $ V \simeq  V_2 + 0.34$ 
 for 
 $V_2 < \V2c ( = 1.32)$.  
 Since the ground state  of  $V_2=0.8 (=V/1.2)$ 
   is the same as that of  $V = V_2 = 0$, 
 the dotted curve of $V_2=0.8$  is the same as 
  that of $V=0$ in Fig.~4 (and  also $ U =4$  in Fig.~3).
 However the solid curve deviates    from that of $V=0$  
  due to $\sigma_c(\omega)$, which depends on 
  not only $U$ but also    $V$ and $V_2$.
  The peaks correspond to the first, third and fifth poles of
 the collective mode where their five poles  located at 
 $\omega$ = 
   0.29, 2.24, 3.03, 3.88 and  4.72
    are slightly different from those of Fig.~3 with $U =4$.   
   Such a  small change of $\sigma (\omega)$ is due to   
    the effect of $V_2$  on  the  charge fluctuation.  
  For $V_2 =1.4$, there appears five peaks  located at  
   $\omega \simeq $ 
   0.9, 2.7, 3.2, 4.1  and  4.8   
 while 
     Eq.~(\ref{pole}) with $q=0$      gives 
  six  poles at 
$\omega  = $
 0.85, 2.59, 2.71, 3.25, 4.12 and  4.80
   respectively 
     with three  continua 
        $  1.15 < \omega < 2.01$, 
           $2.72 < \omega < 3.23 $  and 
         $ 4.15 < \omega < 4.59 $.
  Thus, the peaks of $\sigma (\omega)$  correspond  
    to the first, third, forth,  fifth and sixth  poles.
   We make   following  remarks as     the effects of large $V_2$.
   The collective mode except for the second pole 
     gives rise to the peak in the conductivity. 
  The continuum of  $\sigma (\omega)$ which exists 
   in the intermediate range of frequency 
     is mainly determined by   
       $\sigma_0(\omega)$,  i.e., that of 
             the single particle excitation. 
 The contribution from the single particle excitation becomes dominant 
   with increasing $V_2$.

%
\begin{figure}[tb]
\begin{center}
 \vspace{2mm}
 \leavevmode
 \epsfxsize=7.2cm\epsfbox{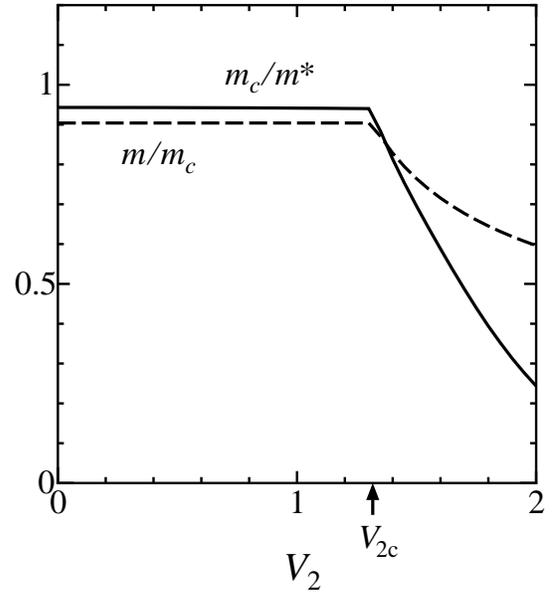}
 \vspace{-3mm}
\caption[]{
 $V_2$ dependence of $m/m_c$ and $m_c/m^*$ 
    for $U=4$,  $V=1.2V_2$ and $\td=0$.   
 The ground state in the region (I) [region (III)]
is obtained   for $ V_2 < \V2c(\simeq 1.32)$  ( $ V > \V2c$). 
}
\end{center}
\end{figure}
%
Using the result of  Fig.~6, we examine 
 the $V_2$ dependence of   $m_c/m^*$ and $m/m_c$, which 
   are  defined by eqs.~(\ref{sum1}) and (\ref{sum3}). 
In Fig.~7, $m_c/m^*$ and $m/m_c$  as the function of $V_2$ 
   are shown by the solid curve  and dashed curve respectively 
   for $U =4$ and $V/V_2=1.2$. 
 It is reasonable that $m/m_c$ is constant for $V < \V2c$,  
 i.e., region (I)
   since the  quantity  given by eq.~(\ref{sum1}) is determined only 
    by the ground state.
 The quantity $m/m_c$ (dashed curve) exhibits  
   a cusp at $V = \V2c$  
    but that of Fig.~5   is continuous at $V=\Vc$.  
 This fact  is comprehended  as follows.    
  The ground state of region (III) is obtained by  a competition 
    between spin density ($S_1$) and charge density ($D_1$) 
  while  that of the region (II) (i.e., the coexistence with $D_2$) 
   is obtained mainly by the  change of the phase  of SDW. 
 It is noticed  that    $ m_c/m^*$ is independent of $V_2$ 
   for $V_2 < \V2c$ 
 although  the pole of   the collective mode depends on $V_2$.
 The rapid increase of $m^*$ for $V_2 > \V2c$ shows a fact that  
   the charge ordering ($D_1 \not= 0)$ induced by  $V_2$ enhances strongly 
  the effective mass of the collective mode.
 For the case of varying $V$ (the solid curve of Fig.~5), 
  such an enhancement takes place for $V$, which is 
   much larger than $\Vc$.  
 Thus it is found that 
     charge ordering of  $D_2$ is less effective compared 
    with that  of $D_1$. 
  
Here we note  the relation between the peak of the conductivity 
 and the pole of the collective mode. 
 As mentioned in the previous section, 
   the collective mode gives rise to  the peak of 
      the conductivity if it is optically active, i.e., 
          the mode can describe the charge fluctuation. 
It is well known that,   in the mode of the incommensurate case, 
  the components of  2$\kf$ SDW and that of  $-2\kf$ SDW are the same 
 in the magnitude but opposite in the sign.
\cite{Lee74}  
Extending the state  to the present case of quarter-filling, 
 we examined eq.~(\ref{pole}) in the limit of small $q$ 
 by choosing  a restricted two-component base  as    
\begin{eqnarray}
 \label{base_m}
 \Phi(q) \rightarrow  
(\e^{-\i\theta}S_{z1}(q)-\e^{\i\theta}S_{z3}(q),D_{2}(q))^t 
\point
\end{eqnarray} 
We have verified that all the pole obtained from  eq.~(\ref{base_m})
 correspond  to the location of the peak of the 
 conductivity  in Figs.~4 and 6.
 For the commensurate case, 
  there are  many poles describing the charge fluctuation 
 while  the pole with lowest energy gives the dominant   weight.

\section{Effective Mass of the Collective Mode}
\setcounter{equation}{0}

 The effective mass $m^*$ of the collective mode has been estimated 
     from the weight of the main peak of the conductivity, 
  $\sigma (\omega)$  in \S 3.
 In this section, we examine the effective mass of
   the collective mode with lowest excitation energy 
    using  a different method  
     and compare with the result of the previous section. 
 For this purpose,  we utilize  the   phase Hamiltonian,  
 which has been  studied for both CDWs
\cite{Fukuyama76}  and SDWs.
\cite{Maki88,TanemuraPTP} 
Using phase variable $\theta(x)$
 which describes the charge fluctuation of SDW with long wave length,
  the effective Hamiltonian is  written as 
\begin{equation} \label{Heff}
\hspace{-2mm}
 H_{\rm eff} = \!\!\int \!\d x \! 
\left[
 \frac{1}{4a} \Pi_{\rho}^2(x) 
+ b \left( \frac{\partial\theta(x)}{\partial x} \right)^2
 + b \, q_0^2 \,\theta^2(x)
\right],
\end{equation} 
where  $\left[\theta(x),\Pi_\rho(x') \right]=\i\delta(x-x')$.
In stead of deriving parameters, $a$, $b$ and $q_0$   in eq.~(\ref{Heff}) 
  from  path integral method,
\cite{Brazovskii} 
 we determine them numerically using the  charge susceptibility and 
 the spectrum of the collective mode 
  [eq.~(\ref{pole})], which has  a gap due to commensurability. 

The coefficient,  $b$, is estimated  as follows.  
From eq.~(\ref{Heff}), the Green function   is calculated  as     
\begin{equation}  \label{green} 
\hspace{-2.1mm}
     \int_0^\beta  \d \tau 
           \left<T_\tau \theta_q(\tau)
           \theta_{-q}(0)   \right>
        \e^{\i \omega_n\tau}
        =   \frac{1/2}{a \omega_n^2 + b (q^2 + q_0^2)} \;,
\end{equation} 
where 
 $\theta_q$ is the Fourier transform of $\theta (x)$. 
It is noted that  the Fourier transform of charge fluctuation with 
 wave number $q$ is given by $\i q \theta_q  / \pi$.   
Then, from eq.~(\ref{green}), 
the static charge susceptibility for small $q$, 
   is written as      
\begin{equation} 
\label{chicp}
  \chi_c(q) \simeq \left(\frac{q}{\pi}\right)^2 
\left< \theta_q \theta_{-q} \right>
=\frac{1}{2\pi^2b}\frac{q^2}{q^2+q_0^2} \point
\end{equation} 
 The charge susceptibility is also calculated 
 from the response function, which is   defined by
\cite{Suzumura_JPSJ95} 
\begin{equation}  \label{response} 
   \frac{-1}{2N} \int_0^\beta \d \tau 
       \left.     \left<T_\tau \Phi(q, \tau)
           \Phi^\dagger(q, 0)   \right>
        \e^{\i \omega_n\tau}
        \right\vert_{\i\omega_n \rightarrow \omega +\i 0} 
                                          \virg 
\end{equation} 
where $\Phi(q)$ is given by eq.~(\ref{base}) and the response function 
 is evaluated in the random phase approximation as 
$
 - \SLP(q,\omega)/
( 1-\SLU(q) \SLP(q,\omega))
$.
 From eq.~(\ref{response}), 
 the static charge susceptibility is also obtained as 
\begin{equation} 
\label{chicRPA}
\chi_c^{\rm RPA}(q) = \left< D_0(q) D_0(-q) \right>
\virg
\end{equation}
\begin{figure}[tb]
\begin{center}
 \vspace{2mm}
 \leavevmode
 \epsfxsize=7.6cm\epsfbox{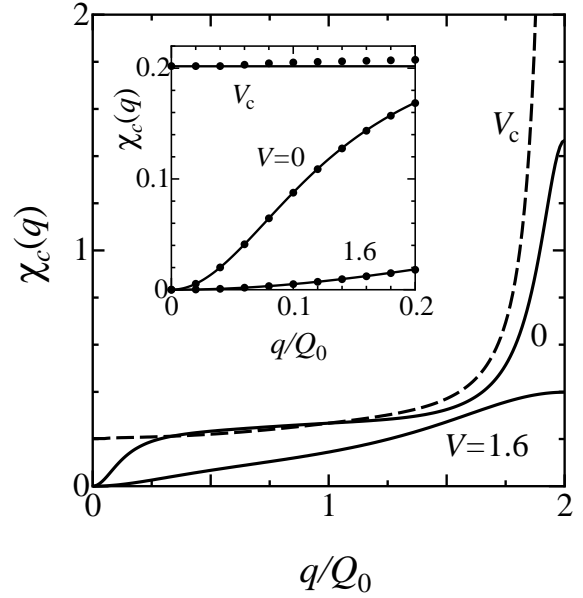}
 \vspace{-3mm}
\caption[]{
 The static charge susceptibility $\chi_c$, as a function of $q/Q_0$, 
for $V=0$, $\Vc(\simeq 0.34)$ and 1.6 where   $U=4$,  $V_2=0$ and $\td=0$. 
In the inset, the numerical results of  $\chi_c(q)$ (dots)  
for the small $q$ are compared with  solid curves obtained 
 from   $\chi_c(q)=q^2[2\pi^2b (q^2+q_0^2)]^{-1}$.
}
\end{center}
\end{figure}
%
 where 
   $D_0(q) = \i q \theta_q / \pi$.
\cite{TanemuraPTP} 
Thus the quantity $b$ can be obtained  by comparing eq.~(\ref{chicp})
 and eq.~(\ref{chicRPA}).
In Fig.~8, 
$\chi_c(q)$ as the function of $q/Q_0$ is shown for 
 $V =0$, $\Vc$ (dotted curve) and 1.6. 
 One finds $\chi_c(0) =0$  except for $V = \Vc$  which leads to  
   $\chi_c(q) \not= 0 $ for $q \rightarrow 0$. 
 The results for $V = \Vc$ is  reasonable 
 if we note   eq.~(\ref{chicp}) does hold and  
  the collective mode becomes gapless, i.e., $q_0 = 0$.  
In the inset, $\chi_c(q)$  obtained from  
 eq.~(\ref{chicp}) (solid curve) is compared with 
 that of eq.~(\ref{chicRPA}) (dots) 
 where    $b$ in eq.~(\ref{chicp})
   has been determined to fit the dots. 
The quantity $a$ is calculated by substituting $b$ into 
   the excitation spectrum 
 obtained from    eq.~(\ref{pole}), which can be  written as 
$\omega_c(q) =  \sqrt{b(q^2 + q_0^2)/a}$ 
 for small $q$.
 In the following calculation, we use renormalized quantities
 defined as  
$\tilde{a} \equiv 4\pi\vf a $ and $\tilde{b} \equiv 4\pi b/\vf$ 
where $\vf=\sqrt{2}$.
 It is found that $\tilde{a} = \tilde{b}=1$ in the limit of weak 
 interaction.  
We note that  $\tilde{a} = m^*/m$ 
 since the first term of eq.~(\ref{Heff}) may be rewritten as 
 $\pi \vf (m/m^*) \Pi_\rho^2$ 
 in a way similar to CDW.
\cite{Fukuyama76} 
Thus the quantity $\tilde{a}$ can be compared with the effective mass 
  of the previous section, i.e., eq.~(\ref{sum3}). 

%
\begin{figure}[tb]
\begin{center}
 \vspace{2mm}
 \leavevmode
 \epsfxsize=6.8cm\epsfbox{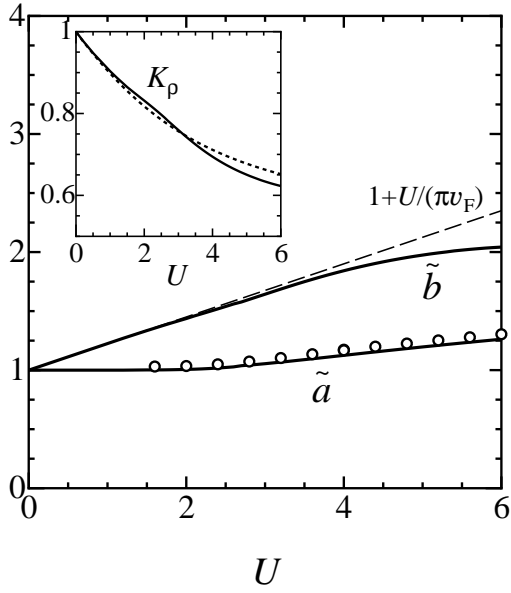}
 \vspace{-3mm}
\caption[]{
 $U$ dependence of $\tilde{a}(=4\pi\vf a)$ and $\tilde{b}(=4\pi b/\vf)$ 
for $V=V_2=\td=0$ where $a$ and $b$ are coefficients of the effective 
Hamiltonian eq.~(\ref{Heff}).  
 The dashed line denotes $1+U/(\pi\vf)$. 
The open circles represent $m^*/m$ estimated from eq.~(\ref{sum3}).  
In the inset,  the solid curve and the dotted curve are 
  $K_\rho(= 1/(\tilde{a}\tilde{b})^{1/2})$ and  
    the exact one, respectively.  
\cite{Schulz_int}
}
\end{center}
\end{figure}
%
%
\begin{figure}[tb]
\begin{center}
 \vspace{2mm}
 \leavevmode
 \epsfxsize=6.8cm\epsfbox{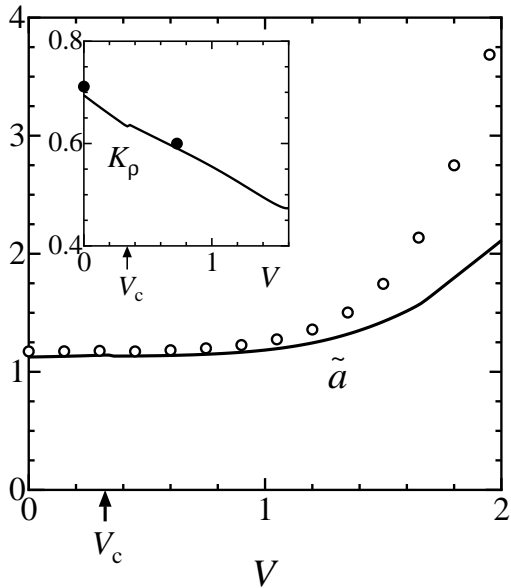}
 \vspace{-3mm}
\caption[]{
 $V$ dependence of $\tilde{a}$ (solid curve) for $U=4$ and $V_2=\td=0$. 
The open circles represent $m^*/m$ obtained from eq.~(\ref{sum3}). 
 In the  inset, solid curve and closed circle  denote   $K_\rho$  and 
  the exact result.
\cite{Mila93}
}
\end{center}
\end{figure}
%
In Fig.~9, quantities $\tilde{a}$ and $\tilde{b}$ 
  as the function of $U$ are shown by the solid curve. 
  The open circle  is  obtained from eq.~(\ref{sum3}) 
 where $\eta = 0.02$ for $U > 4$ 
   and  $\eta = 0.02 ( E_2(Q_0/2)-E_1(Q_0/2))$ otherwise  
    for numerical accuracy. 
 The open circle coincides well with $\tilde{a}$.  
 There is  the visible  enhancement of the effective mass  $m^*$ 
 for $U \gsim 4$.    
The $U$ dependence of  $\tilde{b}$ is also shown by the solid curve.   
For comparison with $\tilde{b}$, we show  
$  1 + U/ (\pi\vf)$ by the dashed line, 
 which is the formula   
  in  the limit of small $U$ and   $V=V_2=0$. 
The  solid  curve deviates from the dashed  line for large $U$ 
 due to the effect of commensurability.  
The good coincidence of $\tilde{b}$ and the dashed line 
 for small $U$ indicates that 
 eq.~(\ref{Heff}) describes well  the effective Hamiltonian 
  as seen in one-dimensional Hubbard model.
\cite{Yoshioka}   
In the inset,  we show   
 $K_\rho(= 1/(\tilde{a}\tilde{b})^{1/2})$, 
   a parameter for the charge fluctuation, by the solid curve 
 where  the dashed curve denotes exact one 
  for one-dimensional Hubbard model.
\cite{Schulz_int}  
The result indicates a fact that 
 the charge  fluctuation around the mean-filed ground state
     may be justified even quantitatively. 

Further we examine $V$ and $V_2$ dependence of  $\tilde{a}$.
In Fig.~10,  the normalized effective mass,  $m^*/m$, 
  as the function of $V$ is shown by the solid curve 
   where the circle denotes the result obtained by eq.~(\ref{sum3}).
 Both results remain  almost the same and independent of $V$ 
for $V \lsim 1$.
 The mass enhancement is found  for $V (\gsim 1)$ which is  
  much larger than $\Vc$. The  enhancement of $m^*$ 
    obtained from the response function is smaller than 
       that obtained from the optical conductivity. 
In the inset, $V$ dependence of 
  $K_{\rho}$ is shown by the solid curve 
    where  the closed circle denotes the  exact result.
\cite{Mila93} 
It is found that  the charge fluctuation 
   obtained by  the present  approximation
    may be  justified also in the presence of $V$. 

%
\begin{figure}[tb]
\begin{center}
 \vspace{2mm}
 \leavevmode
 \epsfxsize=6.8cm\epsfbox{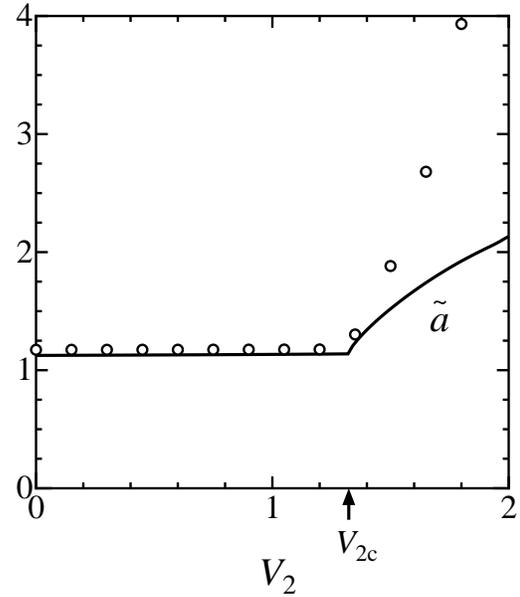}
 \vspace{-3mm}
\caption[]{
 $V_2$ dependence of $\tilde{a}$ for $U=4$, $V=1.2V_2$ and $\td=0$. 
The open circles represent $m^*/m$ obtained from eq.~(\ref{sum3}).
}
\end{center}
\end{figure}
%
In Fig.~11,  $V_2$ dependence of  the normalized quantity 
 $m^*/m$  is  shown 
 where open circle denotes the result obtained by eq.~(\ref{sum3}).
Both are almost the same and stay  constant for $V_2 <  \V2c$. 
In contrast to Fig.~10, $m^*/m$ increases rapidly for 
 $V_2 > \V2c$. 
 The effective mass obtained from eq.~(\ref{sum3}) 
   is larger than that of the present section. 
 The former is the direct calculation while the latter is the result 
 under the   assumption of phase Hamiltonian of eq.~(\ref{Heff}).
 The reason for the discrepancy  for large $V$ and $V_2$ 
 remains unclear at present.

\section{Dielectric Constant and Reflectivity}
\setcounter{equation}{0}

In order to examine the reflectivity, we calculate  
 the dielectric constant $\epsilon (\omega)$, which   is given  by 
\begin{equation}
\label{dielectric}
 \epsilon(\omega) = 1 + \frac{4 \pi \i}{\omega}
      \left( \sigma(\omega) + \i \sigma'(\omega) \right)  \virg 
\end{equation} 
 where   $\sigma(\omega) + \i \sigma'(\omega)$ is  
  the dynamical conductivity defined  by eq.~(\ref{sigmadef}). 
 The imaginary part,  $\sigma'(\omega)$, 
  can be calculated using   the Kramers-Kronig 
  relation with the optical conductivity $\sigma(\omega)$. 
However we calculate $\sigma'(\omega)$ directly   
  from  eqs.~(\ref{sigma0}) and (\ref{sigmac}). 
 Actually,   one obtains 
  $\sigma'(\omega)  =   \sigma_0'(\omega) +  \sigma_c'(\omega)$, 
 where 
\begin{eqnarray}
\label{sigma0m}
 \sigma_0' (\omega) &=& {\rm Im} \left\{ \frac{2}{\i\omega} 
\left[ \Pi_{jj}^0(\omega) - \Pi_{jj}^0(0) \right] \right\} \virg 
\\
\label{sigmacm}
 \sigma_c' (\omega) &=& {\rm Im} \left\{ \frac{2}{\i\omega} 
             \Pi_{jj}^c(\omega) \right\} \point
\end{eqnarray}

In Fig.~12, $\sigma'(\omega)/e^2$ and 
    ${\rm Re}[ \epsilon (\omega)]$ are  shown 
 for $U=4$ and $V=V_2=\td=0$. 
 The quantity $\sigma'(\omega)/e^2$  is 
   the odd function with respect to $\omega$. 
 One finds that    $\sigma'(\omega)/e^2 \propto - \omega$
   for small $\omega$ and changes the sign at $\omega = \omega_c(0)$.
  With increasing $\omega$,
  ${\rm Re}[ \epsilon (\omega)]$ increases rapidly  
  but decreases to change the sign at the frequency, $\omega_c(0)$,  
 corresponding to  
   the main  peak of $\sigma (\omega)$. 
 The lower panel  at $\omega =0$ reads  
   ${\rm Re}[ \epsilon_0 (0)]$ = 1 (dotted curve)
   and ${\rm Re}[ \epsilon (0)]  \simeq 62$ (solid curve).
 Thus, the effect of the collective mode is noticeable 
 even at low frequency. 
 This fact can be comprehended  by  the phase Hamiltonian 
 derived in \S4.
 From eq.~(\ref{green}),
  the  conductivity is calculated  as 
\cite{Fukuyama76} 
\begin{equation}
 \label{phase_conduct} 
 \sigma(\omega) + \i \sigma'(\omega) = - \i \omega \frac{e^2}{\pi^2} 
  {\cal D} (q=0, \i \omega_n)
\bigr|_{\i \omega_n \rightarrow \omega + \i \eta}
 \virg
\end{equation} 
where $ {\cal D}(q, \i \omega_n)$ is the Green function defined  
 by the r.h.s. of eq.~(\ref{green}). 
Substituting eq.~(\ref{phase_conduct}) into eq.~(\ref{dielectric}),
  the dielectric constant    is estimated  as 
\begin{equation}
\label{epsilon}
  \epsilon (\omega)  \simeq          
         1 - \frac{\omega_{\rm p}^2 m_c/m^*}
{(\omega + \i \eta)^2 - \omega_c^2(0) } 
                    \virg 
 \end{equation} 
where $\omega_c(0)(  =q_0\sqrt{b/a})$ 
 corresponds to  the pole of the collective mode with lowest energy. 
The quantity $\omega_{\rm p}$ denotes a plasma frequency defined by 
\begin{equation} \label{plasma}
  \int_0^\infty \d\omega \:\sigma(\omega) = 
   \frac{\omega_{\rm p}^2}{8}
\point  
\end{equation} 
It is found that  eq.~(\ref{epsilon}) 
   explains the behavior  of
    ${\rm Re}[ \epsilon (\omega)]$. 
We note that 
$\epsilon (\omega)=1- (\omega_{\rm p}^2 m_c/m) /\omega^2$  
in the limit of large $\omega$ where 
$\omega_{\rm p}^2 m_c/m = 8\sqrt{2} \,e^2$. 

In terms of the complex dielectric constant given by 
 eq.~(\ref{dielectric}),  the reflectivity is calculated  as 
\begin{equation}
\label{reflectivity}
  R(\omega) = \left|\frac{\sqrt{\epsilon(\omega)} -1} 
                         {\sqrt{\epsilon(\omega)} + 1} 
            \right|^2
                     \point
\end{equation} 

%
\begin{figure}[tb]
\begin{center}
 \vspace{2mm}
 \leavevmode
 \epsfxsize=7.8cm\epsfbox{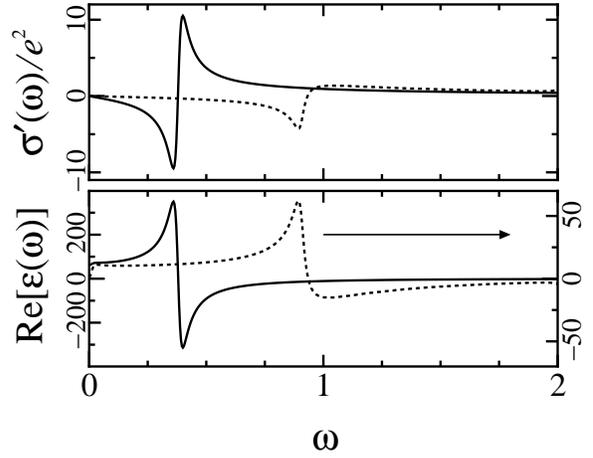}
 \vspace{-3mm}
\caption[]{
 $\omega$ dependence of 
 the imaginary part of the conductivity [$\sigma'(\omega)$] and 
 the real part of the dielectric constant [$\epsilon (\omega)$] 
  for $U=4$, $V= V_2 =0$ and $\td=0$ 
  where the solid curve (dotted curve) has been evaluated 
      from  $ \sigma'$ 
        ($\sigma'_0$).
}
\end{center}
\end{figure}
%
%
\begin{figure}[tb]
\begin{center}
 \vspace{2mm}
 \leavevmode
 \epsfxsize=7.8cm\epsfbox{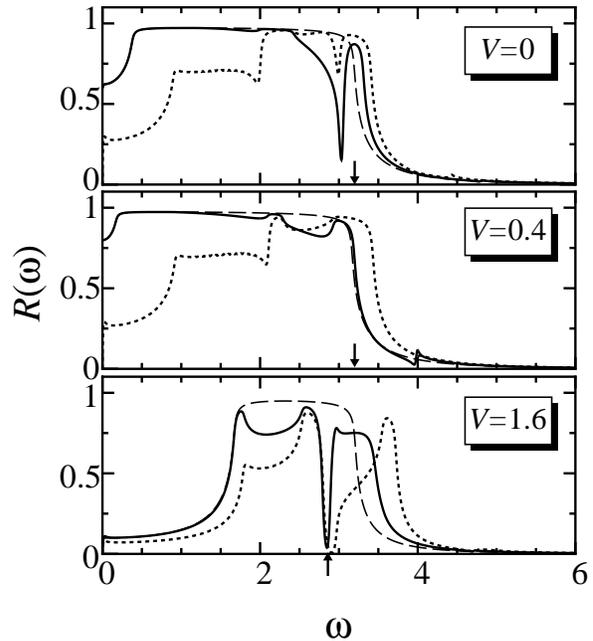}
 \vspace{-3mm}
\caption[]{
 $\omega$ dependence of reflectivity $R(\omega)$ 
  for $U=4$, $V_2 =0$ and $\td=0$  
 with the fixed $V=0$, 0.4 and 1.6 
  where the solid curve (dotted curve) has been evaluated 
       from $\sigma + \i \sigma'$ 
        ($\sigma_0 + \i \sigma'_0$).   
 The arrow denotes a plasma frequency obtained from eq.~(\ref{plasma}). 
The dashed curve is calculated from eqs.~(\ref{epsilon}) 
  and (\ref{reflectivity}).
}
\end{center}
\end{figure}
%
In Fig.~13, $R(\omega)$ is shown for 
 $U=4$, $V_2=\td=0$ and fixed $V=0$, 0.4 and 1.6 
 where the solid curve is obtained from the conductivity 
 with the collective mode 
    and the dotted curve is obtained from the conductivity 
     without the collective mode.  
The arrow denotes  the location of  the  plasma frequency 
  defined by  eq.~(\ref{plasma})   
 where we take  $e=1$. 
The upper panel corresponds to the result for the conventional 
  Hubbard model  with the intermediate interaction.
 The  dotted  curve exhibits   the  insulating behavior
  (i.e., small $R(\omega)$)
 not only  below the gap but also in a certain region  above the gap 
 where the cusp of the dotted curve at $\omega \simeq 0.9$ 
 corresponds to 
  the gap for  the single particle excitation.  
 The solid curve shows that the collective mode 
 enhances the reflectivity in a wide region above the gap of the 
  collective mode [$\omega_c(0)$].
 This comes from the huge magnitude of the dielectric constant  
  as seen from Fig.~12 and eq.~(\ref{epsilon}).
 Here we note  the following fact indicating a significant  role  
 of the collective mode with  $\omega_c(0)$.
  The dashed curve  in Fig.~13 
       is calculated from eqs.~(\ref{epsilon}) 
  and (\ref{reflectivity}). One finds 
 a  good  coincidence between the solid curve and the dashed curve 
  in the wide frequency region except for high frequency regime where 
   the effect of commensurability becomes complicated. 
 Thus the  reflectivity in the most region of frequency is determined by 
  the collective mode with the lowest excitation energy.  
   The sharp dip just below the plasma frequency  
   is found for the frequency close to 
      the second peak of the conductivity (first panel of 
    Fig.~4). 
   The dip at $\omega = 3.03$ originates in the fact that 
    ${\rm Re} [\epsilon (\omega)]$ takes a local maximum  
    ($\simeq 0.13$)   at $\omega = 3.03$  and  becomes negative 
    for $ \omega <3.00$ and  $ 3.04 < \omega$. 
The effect of third peak of the conductivity is invisible 
 in the reflectivity.
  The collective mode enhances the reflectivity 
 at low frequency but suppresses  it at high frequency.
We note that the solid curve becomes the same as that 
 of the free particle (i.e., the metallic state) 
 in the limit of small $U$, 
 which leads to the conductivity $\propto 1/({\rm i}\omega - \eta)$.

%
\begin{figure}[tb]
\begin{center}
 \vspace{2mm}
 \leavevmode
 \epsfxsize=7.8cm\epsfbox{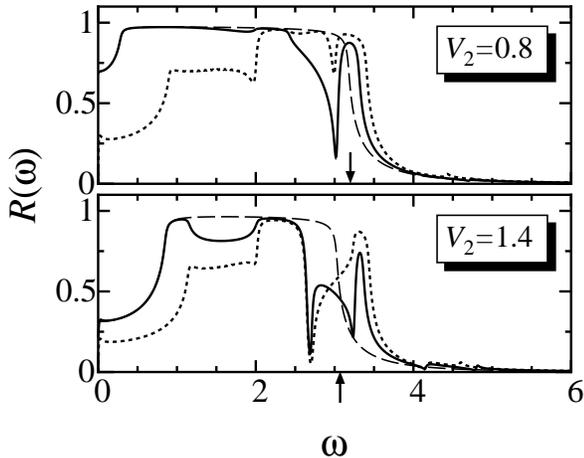}
 \vspace{-3mm}
\caption[]{
 $\omega$ dependence of reflectivity $R(\omega)$ 
  for $U=4$, $V=1.2V_2$ and $\td=0$  
 with the fixed $V_2= 0.8$ and 1.4  
  where  the notations are the same as Fig.~13.
}
\end{center}
\end{figure}
%
Now we examine the effect of inter-site repulsive interaction 
  on $R(\omega)$.   
 The middle panel of Fig.~13
  shows the behavior for $V$ just above $\Vc$ 
 where  the charge ordering begins to develop.
 Compared with the upper panel,   $R(\omega)$ is enhanced 
 at low frequency  due  to the decrease of the gap 
  of the collective mode. 
 The small  dip just below $\omega_{\rm p}$ 
   originates in the single particle contribution while 
    the sharp dip at $\omega \simeq 4$ comes from the 
     collective mode. 
We note the special case of  
   $V=\Vc$ which leads to the gapless excitation of the collective mode.  
 In this case, $R(\omega)$  is expected to be 
    similar to that of free particle 
     except for the region at high frequency just below the plasma 
       frequency    as seen in the middle panel.   
 The lower  panel shows the behavior for $V$ much larger than  $\Vc$
 where  the charge ordering  develops well.
 We find the insulating behavior of $R(\omega)$  
  in the wide range of low frequency 
 and still  $R(\omega) < 1 $ above the gap.
 Thus,  $R(\omega)$ exhibits   
   the insulating behavior for all the frequencies 
 when the charge ordering well develops to 
   localize the electron.  
Since the difference between the solid curve and 
 the dashed curve increases with increasing $V$,
 the charge ordering reduces the effect 
 the collective mode with $\omega_c(0)$. 
In Fig.~14, $R(\omega)$ is shown for 
 $U=4$, $V/V_2=1.2$ and $\td=0$, and fixed $V_2 = 0.8$ and 1.4  
 where the notations are the same  as Fig.~13. 
The upper panel is similar to that of $V=0$ of Fig.~13 
 since both ground states  are in the region (I). 
The lower panel is the result for the ground state in the region (III)
 where  charge ordering comes from $V_2$.     
 The insulating behavior is also found but is slightly complicated 
 at high frequency. 
 Two kinds of dip are located close to the second and third peak 
  of the conductivity (lower panel of Fig.~6) 
   where the origin is the same as that of Fig.~13. 
Based on Figs.~13 and 14, it is found that the reflectivity 
  is suppressed sufficiently by  charge ordering.

\section{Discussion}
\setcounter{equation}{0}

We have examined the effect of the collective mode on 
 the  optical conductivity  and the reflectivity 
  for SDW with one-dimensional quarter-filled band
    using the extended Hubbard with intermediate coupling constant.
 The collective mode with lowest excitation energy 
  exhibits a dominant contribution 
    not only for the limit of small interactions
     but also for the intermediate coupling. 
 When  $U$ increases to the order of band width,
   in addition to the main peak of the conductivity,  
    other peaks become visible  at high frequencies, 
     which correspond to  poles of the mode being active to 
       the charge response. 
 The effect of  the collective mode decreases in the presence of 
   the inter-site interaction, which induces charge ordering.
 The effective mass estimated from the weight of  
  the main peak of the conductivity  increases 
   when the charge ordering is induced by inter-site repulsive interactions 
            being larger than the critical value. 
 The enhancement of effective mass is also analyzed in terms of 
 the phase Hamiltonian. Both results agree well for weak interactions 
 while the latter seems to underestimate in case of 
    strong interactions.     
 The effect of the collective mode on the  reflectivity is 
  further examined. 
We found that  the  mode with lowest excitation  energy determines 
  the reflectivity in a wide range of frequency 
    between the single particle gap and the plasma frequency. 
  The reflectivity in the  range remains large 
    for SDW without  charge ordering while  
      it is reduced in the presence of charge ordering. 

%
\begin{figure}[tb]
\begin{center}
 \vspace{2mm}
 \leavevmode
 \epsfxsize=7.8cm\epsfbox{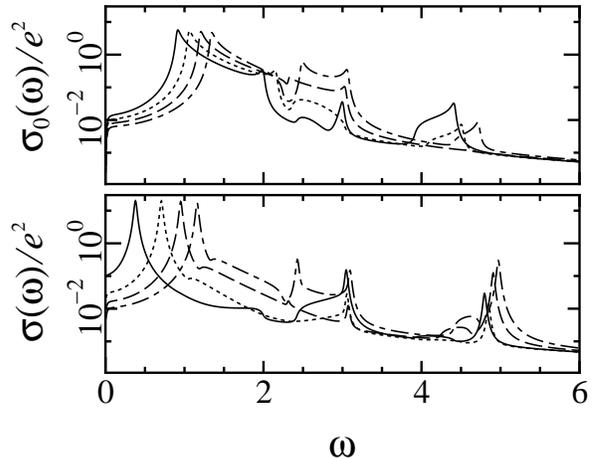}
 \vspace{-3mm}
\caption[]{
Optical conductivity  $\sigma (\omega)$ is shown by the solid curve
for $\td= 0$ (solid line), 0.08 (dotted line), 0.16 (dashed line) 
 and  0.24 (dash-dotted line) 
 with $U=4$, $V = V_2 = 0$  and $\eta=0.02$ 
 where the notations are the same as Fig.~3.
}
\end{center}
\end{figure}
%
%
\begin{figure}[tb]
\begin{center}
 \vspace{2mm}
 \leavevmode
 \epsfxsize=7.cm\epsfbox{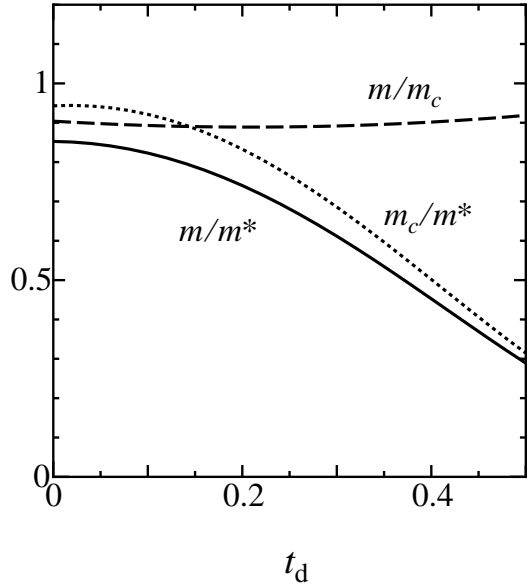}
 \vspace{-3mm}
\caption[]{
 $\td$ dependence of $m/m^*$ (solid curve), 
   $m_c/m^*$ (dotted curve) and $m/m_c$ (dashed curve)
    for $U=4$,  $V= V_2=0$ and $\eta=0.02$.   
}
\end{center}
\end{figure}
%
Here we examine  the effect of dimerization 
 which  is introduced   in  eq.~({\ref{H0}}).   
The main effect of $\td$ on the  ground state is to extend 
 the region (I) into region (II) (the region (III) is less effected)
\cite{TomioLET} 
 since the dimerization is compatible with the spatial 
   variation of spin density of the region (I). 
The effect on the charge gap of the collective mode is rather complicated.
 The charge gap vanishes at   
  the  critical value, $\Vc$ corresponding 
   to the onset of charge ordering, 
  which    increases  with increasing $\td$.
\cite{Suzumura97}  
 Thus $\td$ enhances  the charge gap  for small $V (< \Vc)$ while 
  $\td$ suppresses the charge gap for large $V$.
 In the rigorous treatment of one-dimensional 
 Hubbard model with dimerization,
\cite{Favand}
the dimerization is crucial to 
 obtain  the main (coherent) peak. 
 In this case,   the location of the peak  is 
  shifted from zero frequency to finite frequency 
   since   the charge gap is produced only by dimerization.
In the present case where  SDW forms the long range order, 
  the charge gap does exist  without dimerization      
     indicating a  quantitative  effect of $\td$ on the coherent part. 
However  one may still expect a noticeable  effect 
   on the continuum (incoherent) part  at intermediate frequency 
   as does for the pure one-dimensional  case.
\cite{Favand}
Actually we obtained the following results. 
Figure~15 shows the  conductivity with $\td =$  0, 0.08, 0.16 and 0.24 
 for  $U = 4$  and $V=V_2=0$  
 where the upper panel denotes $\sigma_0$ 
 and the lower one denotes $\sigma$.   
 With increasing $\td$,  
   $\sigma_0$ moves to the regime of high frequency. 
   For $\sigma$, one finds several characteristics.   
 The weight of the first peak decreases as the function of $\td$
  and a  shoulder just above the first peak appears.
   The second peak decreases (increases) for $\td <0.16$ ($0.16< \td$)
  and  
  the third peak decreases (increases) for $\td <0.08$ ($0.08 < \td$) 
   with increasing $\td$. 
  Another peak appears just below the second peak.  
 These come from the interplay of the excitation of the collective mode 
  and that of the single particle. 
 In Fig.~16, the effective mass is shown as the function of $\td$.
The total weight of the conductivity  
 ($m_c/m$) remains almost constant or even  slight decreasing 
  as the function of   $\td$ 
 while the effective mass ($m^*/m$) determined by the first pole 
   is increased  by $\td$.  

Finally, we comment on the relevance of the present result 
 to  the optical experiment 
 in the materials with the one-dimensional quarter-filled band.   
 Although the experiment exhibits 
  various kinds of additional effects such as the coupling to phonon 
 and  one-dimensional fluctuation in the presence of 
 the correlation gap, it is expected that, in  organic conductors,  
 the small reflectivity of  TMTTF salt  compared with 
  that of   TMTSF salt  
  could be due to the  presence of charge ordering.
\cite{Gruner_Science} 

\section*{ Acknowledgments }
 We thank  S. Sugai for useful discussion on the optical conductivity 
 and the reflectivity. 


\end{document}